# Magnetic Measurements of HL-LHC AUP Cryo-Assemblies at Fermilab

J. DiMarco, P. Akella, G. Ambrosio, D. Assell, M. Baldini, G. Chlachidze, S. Feher, J. Nogiec,
V. Nikolic, S. Stoynev, T. Strauss, M. Tartaglia, P. Thompson, D. Walbridge
*Fermi National Accelerator Laboratory*
Batavia, IL, USA

W. Ghiorso, X. Wang
*Lawrence Berkeley National Lab*
Berkeley, CA, USA

*Abstract*— LQXFA/B production series cryogenic assemblies are being built for the LHC upgrade by the HL-LHC Accelerator Upgrade Project (AUP). These contain a pair of MQXFA quadrupole magnets combined as a cold mass within a vacuum vessel, and are to be installed in the IR regions of the LHC. The LQXFA/B are being tested at 1.9 K to assess alignment and magnetic performance at Fermilab's horizontal test facility. The ~10 m - long assembly must meet stringent specifications for quadrupole strength and harmonic field integrals determination, magnetic axis location, and for variations in axis position and local field profile. A multi-probe, PCB-based rotating coil and Single Stretched Wire system are employed for these measurements. To accurately determine rotating coil location and angles within the cold mass, a laser tracker is utilized to record multiple targets at one end of the probe. This paper describes the measurements, probes/equipment, and techniques used to perform the necessary characterization of the cold mass.

## I. Introduction

The US-HiLumi Accelerator Upgrade Project (AUP) being implemented by Fermilab in association with Brookhaven National Lab (BNL) and Lawrence Berkeley National Lab (LBNL) is fabricating 20 4.5 m-long $Nb_3Sn$ superconducting magnets for the interaction regions of the LHC. These magnets are tested vertically at BNL to verify performance specifications, and then pairs of the magnets are combined at Fermilab into a single cold-mass and placed within a cryogenic vessel. This final assembly is then tested horizontally at Fermilab's Magnet Test Facility.

Magnetic measurements are used during fabrication of the cold-mass to verify and help achieve the required alignment between magnets. Additional alignment measurements are made during cold test to be able to report the final magnet alignment parameters to be used for installation at CERN. During the cold testing, measurements are also made to characterize the magnet performance in terms of field quality and strength. This paper discusses the LQXF magnetic measurements being done during fabrication, and during final magnet evaluation under cryogenic conditions.

## II. Requirements

### A. Alignment

The LQXFA must satisfy alignment requirements at operating current at 1.9K [1], and measurements to verify this are part of the cold test regimen. The same requirements are applied during fabrication so that adjustments can be made accordingly. Individual MQXFA magnets for the cold mass have first tests warm at LBNL, including local field offset and field angle determination [2] as an initial quality check. When the MQXFA are brought to the assembly area at Fermilab for the LQXF cold mass fabrication, they are placed on tooling which has been nominally aligned so that the MQXFA should be mechanically within tolerances. The alignment requirements for the two magnets in the cold mass shell are for no point on either magnet to be more than 0.5mm from the average magnetic axis of the two-magnet system, where the average axis is defined as the line going through the midpoint of each individual magnet axis (nodal points). The roll angle between the two magnets is adjusted to be less than 2 mrad, and the cold mass aligned to within 5 mrad with respect to the cryogenic vessel. In addition, during assembly, the axial distance between the magnet centers must be set to within 5 mm.

### B. Strength, uniformity, and harmonic field content

Specification of field quality parameters for the MQXFA are given in [3]. These require measurements of [4]:

- Integrated quadrupole field with accuracy better than 0.1 % and resolution better than 0.01 %
- Harmonic field integrals and magnetic length
- End field harmonics determined with z increments < 0.25 m axially
- Harmonic coefficients with accuracy of 5ppm (10 ppm) of main field in body region for orders 3 to 10 (11 to 14), R = 50 mm
- Local twist variation of the quadrupole roll angle as a function of axial position at the 0.5mrad level
- Local magnetic axis ('waviness') with respect to external fiducials at the 0.2 mm level

## III. Alignment Measurements

A Single Stretched Wire (SSW) system is used to determine average axis and roll angle for both warm fabrication purposes and for final cold alignment measurements. The SSW techniques are effectively those used for LHC IR quad measurements in the early 2000's [5], featuring measurements with AC current on the magnet at room temperature, removal of sag effects, and determination of axis for each magnet individually. However, two new systems have been fabricated with Aerotech stages and software implemented on a modern platform with a scripted framework using LabView code [6] with the newer Metrolab FDI integrator. A first copy of this most recent version of the SSW system was delivered to KEK and was successfully used for in-situ alignment of the Belle II IR [7].

One of the challenging aspects of SSW measurements is the removal of sag effects – which can be as large as 3mm during the cold tests at Fermilab with 18 m wires - and measurement of roll angle, which relies on micron level resolution on measurements made as a function of vertical offset (see [5,6]). First test results of a typical wire vibration frequency determination using the new system at room temperature conditions on the first cold-mass at Fermilab can be seen in the inset of Fig. 1. The frequency dependence at various tensions, is also seen in Fig. 1, showing sag compensation to better than the ~1% level when extrapolated to infinite tension. First results of the magnets with AC current measurements as mounted on the stands show them aligned at a level of ~0.8 mm (Fig. 2). Though this is slightly outside the 0.5 mm tolerance, this is before welding of the stainless steel shell, which will likely improve the alignment. Roll angle measurements repeated at the level of 10 $\mu$rad, indicating determination of offsets at the micron level.

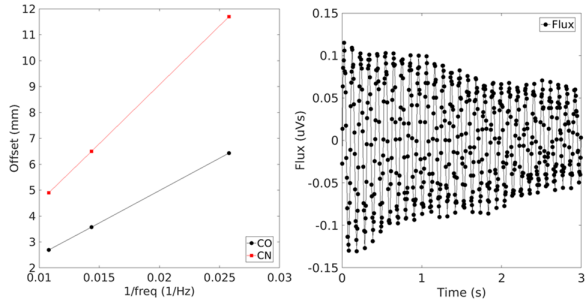

Fig. 1. Frequency measurement: Wire vibration during measurements of LQXFA01 (right), and extrapolation of offsets to $1/f^2 = 0$ after combining results vs various tensions (left).

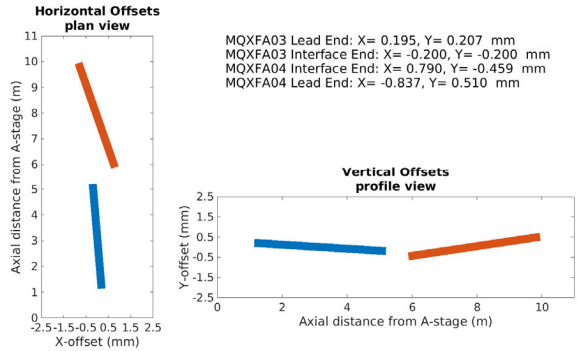

Fig. 2. Results of first alignment measurements of LQXFA01.

Another critical alignment parameter in the cold mass construction involves setting the longitudinal centers of the two magnets the correct distance apart to within a few mm. The SSW system can be used for this measurement as well.

Horizontal transverse 'co-directional' wire movements (stages at magnet ends moving in same direction) and 'counter-directional' movements (stages moving in opposite directions) are used to measure flux changes caused by wire motion, with the wire start position being the magnet axis. Using the known length of the wire between stages, $L_w$, and these flux change measurements, $\varphi_{CO\_Ave}$ and $\varphi_{CN\_Ave}$, as well as the magnetic length, $L_m$, as determined during warm rotating coil measurements during magnet fabrication at LBNL, the longitudinal magnetic center, $Z_c$, can be found from

$$Z_c \equiv Z_a + \frac{L_m}{2} = \frac{L_w}{2} \mp \frac{1}{2*\sqrt{3}}\sqrt{\frac{3\varphi_{CN\_Ave}}{\varphi_{CO\_Ave}}L_w^2 - L_m^2} \quad (1)$$

where $z_a$ is the distance from the end of the wire on the stage to the beginning of the magnetic length. The flux changes can be easily determined to better than 5e-8 Vs, and the wire length to better than 0.1 mm from laser tracker survey. An analysis shows that the dominant source of error stems from uncertainty in $L_m$, which is typically measured at a level of 1 mm. The effect however is not very large, about 0.3mm error in $Z_c$ per mm error in $L_m$, owing to the large wire motions that can be made in the 150 mm magnet aperture. Another source of error is intrinsic to the derivation of (1): namely that for this equation, the magnet strength was assumed to be a step function of length $L_m$. A simulation was performed using a magnet profile that had triangular ends (which closely match warm rotating coil scan data) instead of the step, and the flux was generated and analyzed as in (1). The results showed a 1 mm shift (for the particular magnet/stage geometry on the production bench) from the correct position of the Z-center. Consequently, compensation for this error is included when $Z_c$ is reported.

The SSW will be used for determining these same quantities with the magnet under cryogenic conditions. DC versions of measurements will be used to obtain actual magnet strength at nominal current. AC measurements with 10A will continue to be used for axis measurements after having confirmed that they agree with axis results at full current – this simplifies swapping high current power leads to obtain results.

IV. ROTATING COIL FIELD MEASUREMENTS

The rotating coil measurements will be performed on Test Stand 4 of the Magnet Test Facility at Fermilab. A schematic of the overall layout in the test area for the magnet, stand and probe drive system is shown in Fig. 3. Details of the facility, equipment, measurements, and analysis are discussed below.

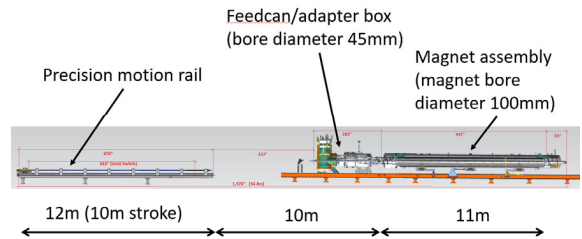

Fig. 3. Overview of test facility magnet layout.

*A. Facility*

The test stand being used is the same as for the first generation interaction region quads built roughly 20 years ago, modified with cryogenic adapter boxes at both ends to receive the 10.2m LQXF magnet. A new 12.5 m -long anti-cryostat 'warm bore tube' (WBT) was fabricated for insertion into the magnet aperture, so that magnetic measurements could be performed with warm probe and magnet at 1.9K. The WBT consists of an inner tube with 104 mm ID/108 mm OD, centered within an outer tube of 123 mm ID/127 mm OD using 3d printed PLA 'spider' standoffs every 0.6 m. The inner tubes are made of 316 stainless steel, and the outer tubes are 304 stainless steel, as 316 SS tubes of required size were not available as stock items. The outer surface of the inner tube also has mounted on it two pairs of 316 stainless, 0.25 mm thick, 12.7 mm wide, heater strips that run the length of the WBT 180 degrees apart from each other. Each heater pair is made by covering one stainless strip completely with Kapton tape as it is secured to the WBT surface, and then laying the second strip precisely on top of it as it is likewise secured. These form a single circuit powered by 1A current with negligible field influence on the measurements and heating capability of 10 W. Fifteen layers of Mylar insulation are wrapped in the annulus between the tubes, and a vacuum port at one end allows this space also to be evacuated when the WBT is being used for measurements. To save cost, the section of the 4.5m-long cryogenic feed box through which the new, large diameter WBT would have had to pass was not rebuilt, but instead an adapter from the old 46 mm ID feedbox WBT to the new WBT was fabricated at the end of the feedbox close to the magnet. A photo of the transition is shown in Fig. 4.

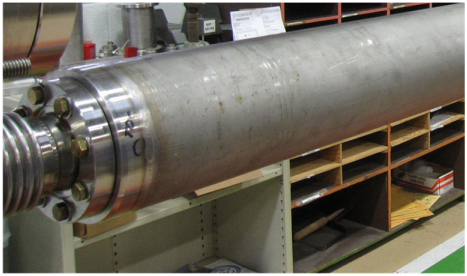

Fig. 4. Transition from 46 mm ID WBT from cryogenic feed-can to 104 mm ID WBT for magnet.

To be able to perform axial scans of the magnet to determine the local strength, harmonics, and twist, the rotating coil will be driven with a thin-walled polycarbonate 'push tube' connected to the carriage of a 10 m long belt-driven rail. Tests of the motion accuracy of the rail showed that the reproducibility between actual and requested positions was quite good, but that linear and offset corrections were needed for both directions of motion. The recorded position of the probe will rely on a laser tracker (LT), which is also being read during the scan measurements. Since at the cryogenic feed end of the magnet the WBT ID is 46 mm, the push tube which connects to the probe for positioning is restricted to this small diameter. To attach the push tube to the probe, the push tube must be inserted all the way through the magnet, and attached outside the far end – the 101.5 mm diameter probe can then be pulled back through the magnet for the scan. Scanning in this direction also keeps a more uniform tension on the push tube, allowing for more reproducible positioning.

The push tube driving the probe contains within it a 6.35mm CF rod used as a quasi-flexible drive shaft for rotating the coil, and is supported within the push tube by small bearings contained in 3d printed holders. The holders also support the electrical cable going to the probe and separate it from interfering with the drive shaft. The ~23 m long drive shaft is rotated at ~4 Hz using a step motor that is connected to the carriage of the belt-driven rail. The carriage and motor housing are designed so that the push tube is always supported over the rail by 3D printed arms with roller bearings on either side (Fig. 5). Thus, no complex moving gate system is needed as the drive carriage travels the length of the rail. The push tube and drive shaft are well-matched in having substantial flexibility, and sag to the bottom of the WBT after about 1m distance both from the probe and from the small to large ID WBT transition. Because of the flexibility and extent of the drive shaft length, sizeable angular speed variations are expected at the coil. But since probe voltages are integrated between angular steps measured by an encoder local to the rotating coil, accurate flux vs angle data can be obtained. The high rotation rate also mitigates against oscillations.

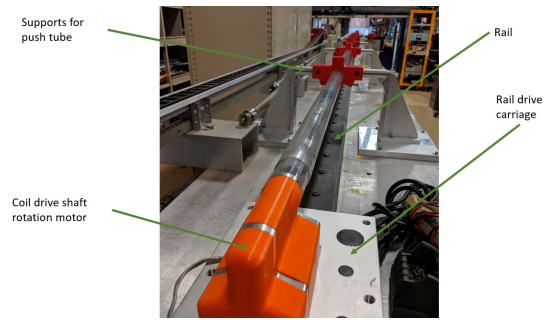

Fig. 5. Rotating coil positioning rail, push-tube, and rotation motor.

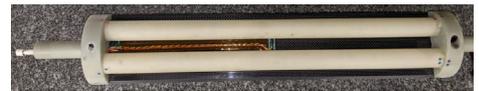

Fig. 8. Rotating coil structure.

*B. Rotating Coil*

The rotating coil has three active probes which simultaneously acquire harmonics data during measurements. The probes differ primarily in length, but are all of similar winding cross-section (Fig. 6). The main probe has an effective length of 436 mm, being a multiple of the 109 mm transposition pitch of the cable. The two other probes have an effective length of 109 mm. The actual winding extent of each probe is 3.6 mm longer than their effective length since the probe windings have 'end regions' where sensitivity decreases. The two 109 mm probes have been fabricated on a four-layer PCB, even though each is a 2-layer design as for the 436 mm probe, in order that they can be overlapped slightly, so that 109 mm movements that match effective length also exactly match the physical position of the probe. This allows the trailing probe of the pair to occupy the same position as the leading probe on the previous move. This is important in the local twist measurements below. Note that though none of the probes have the coil rotation axis going through their center, that the sensitivities are given exactly by

$$K_n = \sum_{j=1}^{N_{wires}} \frac{L_j R}{n} \left( \frac{(x_j + i y_j)}{R} \right)^n (-1)^j \quad (2)$$

(where $j$ sums over every wire of the probe, $L$ are the wire lengths, $x, y$ the positions in the probe cross-section, and $R$ is the reference radius (= 50 mm)), and so the harmonics analysis will correctly compensate.

To fabricate the rotating coil, fixturing was developed for an optical table so that the probe PCBs could be sandwiched between aerospace-grade carbon fiber plates and uniformly epoxied together with high flatness and straightness, forming an extremely stiff three-probe package (Fig. 7). The rotating coil structure for the probes was made from 25mm thick G10 end plugs connected by 30mm G10 tubes (Fig. 8). A precise slot for light press-fit was milled into the end plugs to receive the three-probe assembly, and the active edge was aligned to the 50.75 mm structure radius on a high-tolerance flat surface and secured with epoxy. 3-D printed bearing blocks with full ceramic bearings and 3-point supports (one of them spring-loaded), are attached to the stems of the end-plugs.

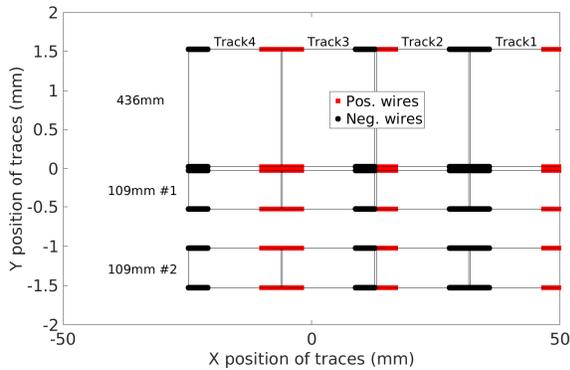

Fig. 6. Cross-section of rotating coil. Red/black wires indicate positive/negative going wires of loops (tracks). The PCB containing the 109 mm probes has a similar thickness to the 436 mm PCB, but uses four layers.

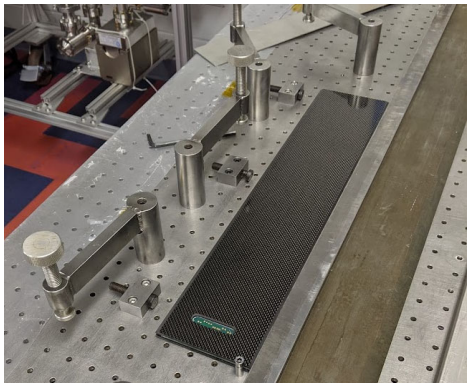

Fig. 7. Fabrication of PCB probe package.

The rotating coil is connected to a detachable 'extension' which contains a local encoder, sliprings, and 3-D accelerometer (Fig. 9). This turns the coil into a nearly self-contained measurement device (similar to 'moles' developed elsewhere [8], [9], except with an external drive here) which we refer to as a FERRET (an acronym for FERmilab Rotating-coil Encapsulated Tesla-probe). The moving portions in the extension are entirely non-metallic to avoid any magnetic effects, especially at the high rotation rate. The slipring has 16 wires and ceramic bearings. The 1024-count encoder was

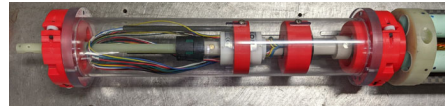

Fig. 9. FERRET extension with encoder, slipring, and 3-D accelerometer.

fabricated at Fermilab using a commercial plastic code wheel and read-head together with acetal bearings. The accelerometer is a MEMS chip, and although it is affected by high magnet field, may provide rough gravity angle or mechanical disturbance information - it is considered accessory, and not at all relied upon. The FERRET extension is at the end of the coil which connects to the push-tube and 10 m drive rail, and mounts to the coil with zero play. Four retro-reflector targets are mounted to the bearing block on the other end of the coil, opposite the drive rail end, and are monitored with a Leica AT960 LT from the far end to give positions and angles of the coil. The targets are on the circumference of a circle with 45 mm radius and are separated one to another in the axial direction by 4mm to make individual targets more distinguishable (Fig. 10). The central target stands 75mm further out to avoid the end of the rotating

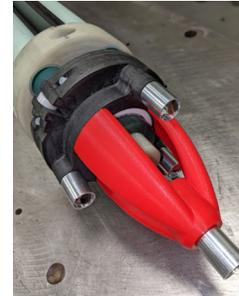

Fig. 10. Laser tracker targets: these are mounted on the bearing block at one end of the coil.

coil shaft.

Each of the three probes of the coil has an unbucked (UB) winding, as well as windings which buck (suppress) dipole (DB), and both dipole and quadrupole (DQB), in a fashion similar to previous probes used for quadrupole measurements [10], [11]. The 9 signals are available on the probe connector, and one of the 109 mm probes shares a common so that the 9 channels can get through the 16 wire slipring and be available at the FERRET connection blocks for selection if needed.

The rotating coil data acquisition system is built on the same EMMA framework that is used for data acquisition with the SSW. The signals are read by 24-bit NI ADC's sampling at 204.8 Hz. Voltage integration is performed within an FPGA, allowing 4 Hz, 8-channel, 1024 samples/rotation data to be continuously displayed and streamed to disk. See [12] for a more complete description.

*C. Measurements*

In terms of the requirements of Section II, axial scans (Z-scans) of the magnets with the rotating coil probe will determine most quantities of interest. In addition, measurements made at single coil positions in the magnet as a function of current will provide data for time and current sensitive variation.

*1) Strength*

Local and end field strength and harmonics will be determined from Z-scans with 109 mm step size. For magnetic length, the contiguous measurements of the 436 mm probe will be used: the positions which integrate through the ends (Lead End and Non-Lead End) will be combined with the body measurements according to

$$L_m = \frac{\int LE + \int NLE + g_{ave\_body} * L_{body}}{g_{ave\_body}} \quad (3)$$

where $L_{body}$ is the distance between coil positions adjacent those that integrated the ends, and $g_{ave\_body}$ is the average gradient inclusive over those positions. Note that the numerator of (3) gives the integrated strength measured by the rotating coil. The integral strength determined with the rotating coil will be cross-checked by similar integral measurements with the SSW, which should be the best measurement in terms of accuracy, owing to the micron level control of wire motions (though at high currents, strength measurements must be made as a function of tension to remove effects of wire susceptibility).

*2) Harmonics*

To achieve the harmonics accuracies needed, each of the three probes will have in-situ calibration using an evaluation of UB vs DB quadrupole (main field) strength. That is, since the DB strength is independent of radial or lateral offset from rotation axis, forcing the UB strength (sensitive to both of those) to be consistent with DB allows determination of those from amplitude and phase comparisons respectively [13]. Note that the calibration procedure can even be applied for every rotation, or conversely used to monitor for changes in the probe over time.

If probe radius were not well known, errors in analysis of harmonics would go approximately as the order times relative error in radius.

$$\frac{\Delta C_n}{C_n} = \frac{n * \Delta r}{r} \quad (4)$$

For example, if the harmonic, $C_n$, has value 5 units, then an error in the harmonic of 0.05 units (requirement) would need relative radius error of $\sim \Delta r = 0.1$ mm for probe radius $r = 50$ mm, with n=5. Since the in-situ calibration achieves better than 0.01mm in determining actual radius, the required accuracy of harmonics wrt this error should be easily achieved.

The other potential source for harmonics error is spurious harmonics from vibrations. This is predominantly mitigated with bucking of main fields, but can stem from other neighboring harmonics as well (e.g. large b6 creating spurious a7). The error from transverse or torsional vibrations goes roughly as

$$\varepsilon \sim n * \frac{\delta}{R} * C_n \quad (5)$$

Where $\delta/R$ is the amplitude of the transverse vibration wrt the reference radius (or similarly the angle error for torsional vibrations (i.e. arc length)), and $C_n$ is some neighboring harmonic of order n which can cause spurious harmonics given the right frequency and amplitude $\delta$. Given the error harmonic, $\varepsilon$, should be less than 0.05 units, and picking some representative values of $C_n = 5\ units$, $R = 50\ mm$, $n=5$, we see that the vibration amplitude needs to be on the order of $\delta = 0.1$ mm to reach this level. The mechanics should be better than this, but other environmental issues will have to be checked. Comparison of two independent coils using the same PCB probes, one used at Fermilab, and the other sent to BNL for use in vertical measurements of the individual MQXFA magnets, were cross-calibrated on model magnet MQXFS1 in June 2017. Results suggest that differences seen are consistent with being able to meet accuracy requirements (Fig. 11), though comparison was with the same data acquisition and rotation system, so differences may not reflect all sources of error during actual measurements.

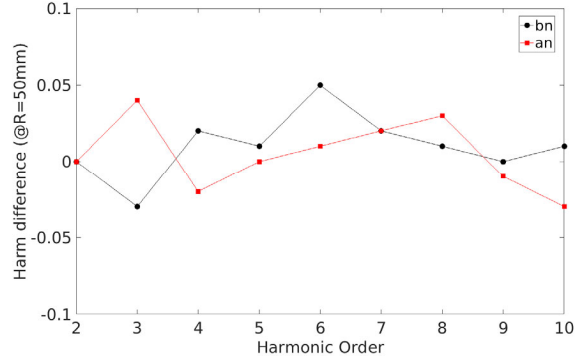

Fig. 11. Comparison of harmonics measured with two probes.

Analysis of the rotating coil data has the option of finding harmonics using traditional FFT determination of flux amplitudes, or a method that determines these from the solution of Eqn. (7). Starting with an expression of probe flux at each of the $k$ angles, $\theta$, being a sum of complex harmonic flux amplitudes $\varphi_n$,

$$\Phi(\theta) = \text{Re}\left[\sum_{n=1}^{\infty}(\varphi_n^{Re} + i\varphi_n^{Im}) * e^{in\theta}\right] \quad (6)$$

we solve a $k \times 2n + 2$ matrix for orders of interest (n < 14 in the case of the LQXFA magnet)

$$\begin{bmatrix} \cos(1\,\theta_1) & -\sin(1\,\theta_1) & \cdots & \cos(n\,\theta_1) & -\sin(n\,\theta_1) & t_1 & 1 \\ \cos(1\,\theta_2) & -\sin(1\,\theta_2) & \cdots & \cos(n\,\theta_2) & -\sin(n\,\theta_2) & t_2 & 1 \\ & & \vdots & & & & \\ \cos(1\,\theta_k) & -\sin(1\,\theta_k) & \cdots & \cos(n\,\theta_k) & -\sin(n\,\theta_k) & t_n & 1 \end{bmatrix} \begin{bmatrix} \varphi_1^{Re} \\ \varphi_1^{Im} \\ \vdots \\ \varphi_n^{Re} \\ \varphi_n^{Im} \\ S \\ O \end{bmatrix} = \begin{bmatrix} \Phi(\theta_1) \\ \Phi(\theta_2) \\ \vdots \\ \Phi(\theta_k) \end{bmatrix}$$

(7)

where the time of the flux sample, $t_n$ (which may be substantially non-uniform owing to the flexible drive shaft), and unity columns allow simultaneous solution of the slope (S) and offset (O) in the flux signal. This form has the advantage of being able to make the best removal of slope from drift based on a single rotation. It also can remove particular noise frequencies if needed; to implement this, columns with $cos(\omega t_n)$, $-\sin(\omega t_n)$ are inserted in (7) to solve (and thereby remove) the amplitudes associated with $\omega$.

This may be useful as the coil will rotate at ~4 Hz, and n = 14 is of interest - so the signal starts getting close to the 60 Hz AC line-cycle frequency. Since the harmonics are limited to frequencies of interest, in general solving (7) is slightly faster than an FFT, except when the number of encoder pulses (*k*) gets large (> ~2000). The redundancy (*k* equations for *2n+2* unknowns) acts as a least squares fit among the samples.

*3) Axis straightness and variation*

Local offset measurements will rely on the probe offset determination with respect to magnetic center, as calculated from harmonics feed-down, coupled with laser tracker measurements which record the physical offset variations. To the extent that probe-measured magnetic center offsets track the physical probe offsets, the magnet axis has no real variation. This axis can also be related to that found with SSW using external fiducials from the magnet and/or building. To calculate the average physical offset of the probe proper based on LT target points at the end of the coil, a Helmert transform is used to perform a seven parameter fit (X, Y, Z, yaw, pitch, roll, scale) which includes 'passive points' representing the center of each probe as determined from a calibration - this provides proper correspondence between target positions and actual offset of probes, including any effects from coil pitch/yaw angles. Repeatability tests in a preliminary arrangement with the WBT suspended across three support points, indicate consistency at better than 100 $\mu$m level even deep within the tube (Fig. 12).

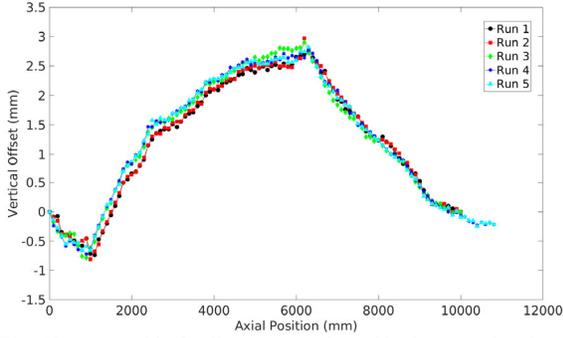

Fig. 12. Centroid of coil targets as viewed by laser tracker from one end of magnet. Note that the first 3 runs are taken about 1 week before the last two, and that the feature at ~ 6000 mm arises because of a joint between two pipes (between the magnets).

*4) Local field angle variation*

To examine local twist variation, the roll as measured by the laser tracker from the probe targets cannot be used, because the bearing block on which the targets reside is not coupled to the angular position of the encoder on the opposite end of the probe. Instead the dual 109 mm – long probes will be used to measure the field angle variation. Consecutive 109 mm steps of the coil place the trailing probe at the location of the leading probe on the previous step, allowing local differences to be measured throughout the scan. Taking the cumulative sum of the measured angle difference between the two probes, with step size defined as $\Delta z \equiv (z_i - z_{i-1})$, gives

$$\Delta\theta(1) = \theta_{magNonLin}(z_1) - \theta_{magNonLin}(z_0) + \beta_{magLin} * \Delta z - \theta_{probeOff}$$
$$\Delta\theta(2) = \theta_{magNonLin}(z_2) - \theta_{magNonLin}(z_0) + 2(\beta_{magLin}\Delta z - \theta_{probeOff})$$
$$\vdots$$
$$\Delta\theta(n) = \theta_{magNonLin}(z_n) - \theta_{magNonLin}(z_0) + n(\beta_{magLin}\Delta z - \theta_{probeOff}) \quad (8)$$

where the $\Delta\theta$ is the local angle variation at each position *n*. The $\Delta\theta$ is comprised of the non-linear variation at each position, $z_n$, with respect to the first position, $z_0$ (which can be subtracted as an offset), plus the overall linear twist $\beta_{magLin}$. Note that the calibrated angle offset between the two probes, $\theta_{probeOff}$, must be subtracted since it is indistinguishable from the linear twist. Taking care of these gives an expression for the local twist along the magnet without needing to have a gravity or external probe angle reference:

$$\Delta\theta(n) = \theta_{magNonLin}(z_n) + n * \beta_{magLin} * \Delta z \quad (9)$$

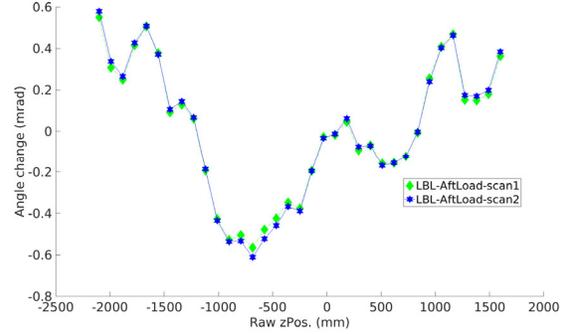

Fig. 13. Determination of local twist using dual 109mm probes on the rotating coil at LBNL for MQXFA04. Repeated scans show average difference of 15 $\mu$rad for a given position.

An example of this technique, plotting $\Delta\theta$ vs $z_n$ during dual probe measurements of a magnet during assembly at LBNL, is shown in Fig. 13. The average repeatability is better than 15 $\mu$rad.

## V. CONCLUSION

Magnetic measurements on production LQXFA have started as the first unit has begun being assembled. Facilities, systems, and techniques have been developed and are in place for the fabrication and cryogenic measurements that will characterize these magnets for use in the interaction regions of the LHC.

ACKNOWLEDGMENT

The authors thank the technical staff of Fermilab's Applied Physics and Superconducting Technology Division for their contribution to these efforts.